\documentclass[page-classic]{epl2} 
\usepackage[T1]{fontenc}
\usepackage[latin1]{inputenc}
\usepackage{graphicx}
\usepackage{dcolumn}
\usepackage{bm}
\usepackage{amsfonts}
\usepackage{epsfig}
\usepackage{mathrsfs}
\usepackage{color}

\def\dd{\mathrm{d}}
\def\g{\mathrm{g}}
\def\gw{\mathrm{gw}}
\def\bcb{\mathrm{BCB}}
\def\cgwb{\mathrm{CGWB}}
\def\la{\leftarrow}
\def\ra{\rightarrow}
\def\lra{\leftrightarrow}
\def\A{\mathrm{A}}
\def\B{\mathrm{B}}

\def\S{\mathrm{S}}

\begin{document}

\title{Large scale EPR correlations and cosmic gravitational waves}
\shorttitle{Large scale EPR correlations}
\author{B. Lamine\inst{1} \and R. Hervé\inst{1} \and M.-T. Jaekel\inst{2}
\and A. Lambrecht\inst{1} \and S. Reynaud\inst{1}}
\shortauthor{B. Lamine \etal}
\institute{
\inst{1} Laboratoire Kastler Brossel - ENS, UPMC, CNRS,
  Campus Jussieu, F-75252 PARIS, France \\
\inst{2} Laboratoire de Physique Th\'eorique - ENS, UPMC, CNRS,
  24 rue Lhomond, F-75231 PARIS, France  }

\abstract{We study how quantum correlations survive at large scales
in spite of their exposition to stochastic backgrounds of
gravitational waves. We consider Einstein-Podolski-Rosen (EPR)
correlations built up on the polarizations of photon pairs and
evaluate how they are affected by the cosmic gravitational wave
background (CGWB). We evaluate the quantum decoherence of the EPR
correlations in terms of a reduction of the violation of the Bell
inequality as written by Clauser, Horne, Shimony and Holt (CHSH). We
show that this decoherence remains small and that EPR correlations
can in principle survive up to the largest cosmic scales. }


\maketitle

It has long ago been suggested by Feynman that gravitation could be
at the origin of a universal decoherence mechanism preventing
macroscopic systems to exhibit quantum coherence
properties~\cite{feynman:lecturesG}. This idea has
been considered since that time through the study of different mechanisms~%
\cite{karolyhazy1966,diosi1989,jaekel1994,penrose1996,kay1998,kok2003},
some of them involving Planck-scale
physics~\cite{percival,amelino2004,wang2006}.

These important theoretical questions can also be addressed as
experimental challenges. In particular the following questions have
been discussed in this context~\cite{arndt2009}~: (\textit{i}) are
quantum interferences limited to microscopic objects having masses
or energies smaller than some intrinsic limit~? (\textit{ii}) are
quantum correlations limited to experiments performed on length
scales smaller than some intrinsic limit~? These two questions have
led to efforts for producing evidence for quantum interferences with
larger and larger molecules~\cite{largemolecules} as well as quantum
correlations on larger and larger distances~\cite{largescaleEPR}. Up
to now, the reported limits are associated with practical rather
than fundamental issues. Efforts are going on for mastering these
technical limitations and approaching closer and closer the
underlying fundamental questions in future
experiments~\cite{arndt2009,ursin2009}.

Here, we will study the effect of the unavoidable interaction with
the gravitational waves backgrounds which pervade our spacetime
environment. It has already been shown that such an interaction was
responsible for an intrinsic decoherence mechanism in atomic
interferometry. The mechanism is more and more efficient for larger
and larger masses, but it depends also on other parameters. With
this model of spacetime fluctuations, it is possible to associate
quantitative estimations to the qualitative Feynman argument, using
only known gravitational physics, astrophysics and
cosmology~\cite{gwdecoh}.

In the present paper, we give quantitative estimations for the case
of Einstein-Podolski-Rosen (EPR) quantum correlations between
polarization entangled pairs of photons~\cite{EPRexpt}. Precisely,
we study how the violation of the Bell inequality~\cite{bell}, under
the form written by Clauser, Horne, Shimony and Holt
(CHSH)~\cite{clauser1969}, is affected after the propagation of
photons over some large distance. We evaluate the effect of the
binary confusion background (BCB) and of the cosmic gravitational
wave background (CGWB)~\cite{gwbackgrounds}. The former is a
classical background resulting from the confusion of gravitational
waves emitted in the Galaxy and its vicinity. The latter, predicted
to have been created from quantum fluctuations of the metric in
primordial cosmology~\cite{grishchuk1975}, is essentially
characterized by the dimensionless parameter $\Omega _\gw$ which
measures the energy density of gravitational waves with respect to
the critical density energy.

The main result of the paper will be that the reduction of the
violation of CHSH inequality is bound by the value of $\Omega _\gw$
and therefore remains small even after propagation at the largest
cosmic distances. This means that EPR polarization entanglement is
predicted to survive at these extreme scales without being washed
out by the interaction with gravitational waves backgrounds.


\section{Polarization entangled photon pairs}

\begin{figure}[h!]
\onefigure{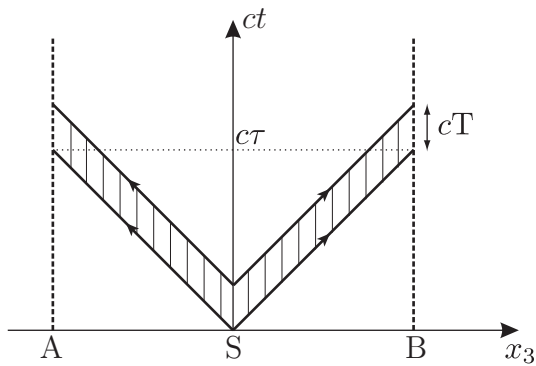}
\caption{Spacetime diagram associated with a Bell test experiment with
polarization entangled photon pairs; the nearly vertical lines represent
the motions of the source S and detectors A and B while the inclined lines
correspond to photons from source to detectors; $\tau$ is the time of flight
from S to A or B and $T$ is the integration time needed to estimate
correlations.}
\label{fig:spacetimeEPR}
\end{figure}

The Bell test experiment considered in this paper is schematized on the
spacetime diagram of fig.~\ref{fig:spacetimeEPR}. A source S (placed for
example on Earth) sends pairs of polarization entangled photons towards two
detectors A and B. The geodesic motion of S, A and B is represented by the
worldlines which are nearly vertical of fig.~\ref{fig:spacetimeEPR} (their
velocities are much smaller than $c$). The lines inclined at $\pm 45^\circ$
with respect to the space and time axis correspond to the propagation of the
photons over a time of flight $\tau$.

The source S emits pairs of polarization entangled photons. We
consider that the corresponding two-photon state has a null momentum
and a null spin (singlet state). From conservation and parity
arguments \cite{feynman:lecturesQM}, it is thus deduced that the
photon pairs are described by a Bell state
\begin{equation}
\vert \psi_\S\rangle=\frac{1}{\sqrt{2}}\left( \vert R_\la\,,\,
R_\ra\rangle - \vert L_\la \,,\, L_\ra\rangle \right)\;,
\label{eq:state}
\end{equation}
where $R$ and $L$ denote right-handed and left-handed circular
polarizations while $\la$ and $\ra$ refer to the photons propagating
from S towards respectively A and B. One of the latter observers,
say A, measures the polarization of photons with polarization beam
splitters and photodetectors. He attributes $+$ and $-$ values to
detections in the two polarization channels, when the polarization
measurement setup has a local orientation defined by the angle
$\Theta_\A$. The origin of this orientation is defined by a local
reference. The observers A and B count the coincidences over a
measurement time $T$ for the 4 possibilities $++$, $+-$, $-+$ and
$--$, and they measure probabilities (average numbers for each
possibility divided by the number of all coincidences)
\begin{eqnarray}
&&P_{++}=P_{--} =\left\vert \frac{e^{i\Theta} - e^{-i\Theta}}2
\right\vert ^2 =\frac {\sin^2 \Theta}2 \;, \nonumber\\
&&P_{+-}=P_{-+} =\left\vert \frac{e^{i\Theta} + e^{-i\Theta}}2
\right\vert ^2 =\frac {\cos^2 \Theta}2 \;, \nonumber\\
&&\Theta=\Theta_\B-\Theta_\A \;. \label{eq:P}
\end{eqnarray}

The two observers A and B then evaluate correlation functions $E$
and $S$ defined respectively for 2 and 4 orientations of the
polarization measurement setups~\cite{clauser1969}
\begin{eqnarray}
&&E^{\A\B}=P_{++}+P_{--}-P_{+-}-P_{-+}=-\cos\left(2\Theta\right) \;, \nonumber\\
&&S=E^{\A\B}-E^{\A\B^\prime}+E^{\A^\prime\B}+E^{\A^\prime\B^\prime}\;.
\label{eq:E}
\end{eqnarray}
Should the conditions of Bell theorem be obeyed~\cite{bell}, the
following CHSH inequality would be deduced~\cite{clauser1969}
\begin{eqnarray}  \label{eq:CHSH}
&& \vert S \vert \leq 2
\end{eqnarray}
As is well known, this inequality is violated by the polarization
entangled photon pairs described by the preceding equations
(1-\ref{eq:E}). The maximum violation is predicted for example with
the angles $\Theta_\A=0$, $\Theta_\B=\pi/8$, $\Theta_A^\prime=\pi/4$
and $\Theta_\B^\prime=3\pi/8$, which lead to $ \vert
S_{\mathrm{max}} \vert =2\sqrt{2}$.

We want to stress at this point that the effects of gravitation have
been ignored in the discussion up to now. In particular, the
orientation of the polarization measurements setups at A and B
raises no fundamental question in the absence of spacetime curvature
but it has to be carefully defined in the context of general
relativity. We note that the orientation can be controlled by using
the expression (\ref{eq:E}), allowing the two remote observers to
exploit the available quantum correlations to get a relation between
the angles $\Theta_\A$ and $\Theta_\B$.


\section{Effect of gravitational fields}


We now consider the same problem in general relativity, taking into
account the effect of gravitational fields, and in particular
gravitational waves, on polarization entangled photon pairs. To this
aim, we have to study the propagation of polarized electromagnetic
fields in general relativity. The dominant effect is a rotation of
the polarization of light along the ray~\cite{polarlightgrav}. The
use of this effect for the purpose of detecting gravitational waves
has been studied~\cite{cruise,tamburini} as well as its possible
contribution to decoherence~\cite{terashima}. Here, we consider the
decoherence of the photon pairs due to the interaction with the
gravitational waves background.

In order to treat consistently this problem in general relativity,
one has to deal not only with the polarization of electromagnetic
field but also with the orientation of the local reference axis at
the two observers. In the eikonal approximation which is sufficient
for the purpose of this letter, both problems are treated by solving
the geodesic equation for the motion and parallel transport for the
polarization of light and orientation of the polarizers. It is only
by taking both effects into account that one obtains a properly
defined observable, associated with a gauge invariant expression. In
the following, we use the results of detailed calculations presented
in~\cite{hervePhD}.

A proper observable is defined by comparing the rotation of the
photon polarization along the ray $\rightarrow$ from the source S to
the detector B and the rotation of local reference orientations at
the source and detector. At first order in the gravitational
perturbation, this observable has the following expression
\begin{equation}
\alpha_\ra(t)=\frac{1}{2}\int_{ct}^{ct+c\tau}(\partial_1
h_{23}-\partial_2h_{13})\mathrm{d}\sigma\;.  \label{eq:rotation}
\end{equation}
The integral is taken along the unperturbed path of the photons from
the time $t$ to the time $t+\tau$, with $\tau$ the time of flight
and $\sigma$ the time parameter along this path. The notations for
the metric are the same as in~\cite{reynaud2009} and the expression
(\ref{eq:rotation}) is written for a propagation along the direction
$x_3$. The angle $\alpha_\ra$ and the similarly defined $\alpha_\la$
are gauge invariant observables.

Considering first the case of stationary gravitational fields, we
see that eqs.(\ref{eq:P},\ref{eq:E}) are changed as follows
\begin{eqnarray}
&&P_{++}=P_{--} =\left\vert \frac{e^{i\Theta}e^{-i\alpha} -
e^{-i\Theta}e^{i\alpha}}2\right\vert ^2 \;, \nonumber\\
&&P_{+-}=P_{-+} =\left\vert \frac{e^{i\Theta}e^{-i\alpha} +
e^{-i\Theta}e^{i\alpha}}2\right\vert ^2 \;, \nonumber\\
&&E^{\A\B}_\g=-\cos\left(2(\Theta-\alpha)\right) \;,
\quad\alpha=\alpha_\ra-\alpha_\la\;. \label{eq:Eg}
\end{eqnarray}
A comparison of (\ref{eq:Eg}) with (\ref{eq:E}) shows that that
stationary gravitational fields lead to a relative rotation of the
orientations at A and B, taking into account the parallel transport
of photon polarization as well as that of orientation of the
polarizers. The structure of the formula can also be understood in
analogy with an interferometric signal~: the rotation of linear
polarizations is indeed equivalent to different dephasings $e^{\pm
i\alpha}$ for the two circular polarization states. Using this
analogy, we will be able in the next sections to use results already
known for interferometers and apply them to the case of Bell test
experiment.

Up to now we have supposed that the angle $\alpha$ is time
independent or at least that is varies slowly enough so that it does
not affect the measurements during the integration time $T$. In this
case, the already discussed trick can still be used~: the two
observers at A and B can in fact compensate for the slow effect of
gravitational fields by adjusting their local orientation
references. It follows that the maximum violation of Bell's
inequality is not affected, though the orientation of the polarizers
may have to be changed in order to attain it. Of course, this can no
longer be the case when rapid fluctuations cause a blurring of the
correlation and lead to a decrease of the maximum violation of CHSH
inequality. We focus the discussion on this effect in the sequel of
this letter.


\section{Effect of stochastic gravitational waves}


We now discuss the effect of stochastic backgrounds of gravitational
waves, which produce not only drifts but also fluctuations.

We first remind that two stochastic backgrounds are usually studied.
The so-called binary confusion background (BCB) is the sum of
signals of all unknown binaries in our Galaxy and its vicinity. Its
stochastic nature is only due to the confusion of a large number of
poorly known sources. In particular, each binary system produces a
classical gravitational wave, so that the superposition of all
contributions remains a classical stochastic background. In
contrast, the cosmological gravitational wave background (CGWB)
stems from quantum fluctuations created during the primordial cosmic
era and then amplified by a huge factor through their coupling to
the evolution of Universe~\cite{grishchuk1975}.

Gravitational waves may be described in linearized general
relativity and the fluctuations of the metric field described by
correlation functions (all notations are as in~\cite{reynaud2009}).
The backgrounds appear stationary for not too long observations. We
also use simple assumptions of unpolarized and isotropic
backgrounds, which are valid for the CGWB and first approximations
for the BCB. All information about the backgrounds is thus contained
in the one-sided spectral density $S_\gw$~\cite{gwbackgrounds}
\begin{equation}  \label{eq:Sgw}
S_\gw[\omega] = \int_{-\infty}^\infty \mathrm{d} t \exp\left(-i\omega
t\right) \langle h(t)h(0)\rangle\;,
\end{equation}
where $h$ is any one of the amplitudes $h_{12}$ or $(h_{11}-h_{22})/2$ (or
other ones obtained after spatial rotations). The spectrum $S_\gw$ can
equivalently be described as a spectral density of energy, or a number of
graviton per mode $n_\gw$ (much larger than unity for all modes of
interest),
\begin{equation}  \label{eq:10}
\hbar\omega n_\gw[\omega]=\frac{5c^5}{16G}\,S_\gw[\omega] \;.
\end{equation}
For the CGWB, the spectrum $S_\gw$ is often written as
\begin{equation}
S_\gw= \frac{12\pi H_0^2}{5\omega^3}\,\Omega_\gw \;,
\label{eq:sigma alpha}
\end{equation}
where $H_0$ is the present day Hubble rate ($H_0=\dot{a}/a$ where
$a$ is the cosmic size parameter) while the dimensionless quantity
$\Omega_\gw$ represents the energy density of the background
compared to the present value of the critical energy density for
closing the Universe.

\begin{figure}[h!]
\onefigure{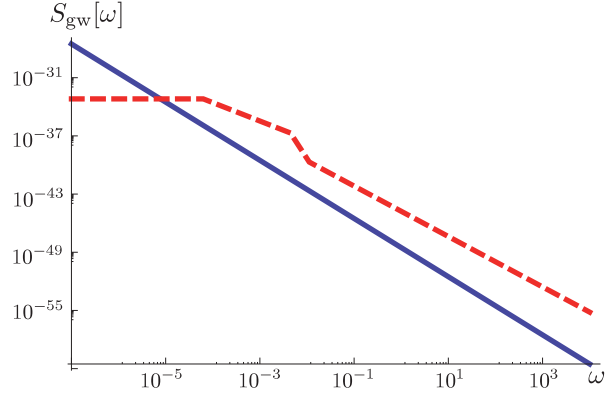} \caption{Spectra $S_\gw$ considered in
the discussions~: the full (blue) line corresponds to the CGWB with
$\Omega_\gw=10^{-14}$ and the dashed (red) line to the BCB (model
given in~\cite{ungarelli2001}). } \label{fig:spectra}
\end{figure}

The parameter $\Omega_\gw$ might depend on frequency but it is often
considered as constant over broad frequency intervals. Though the
CGWB has not been detected up to now, constraints on the value of
$\Omega_\gw$ may be drawn from a variety of
observations~\cite{gwbackgrounds}. In particular, $\Omega_\gw$ has
to be smaller than a few $10^{-14}$ at $\omega\sim10^{-16}$Hz
(cosmic limit), a few $10^{-8}$ at $\omega\sim10^{-8}$Hz (pulsar
limit), and a few $10^{-6}$ at $\omega\sim10^{2}$Hz (laser
interferometry limit)~\cite{abbott2009}. We have drawn as the full
(blue) line on Fig.\ref{fig:spectra} the spectrum $S_\gw$
corresponding to the CGWB with $\Omega_\gw=10^{-14}$. For the sake
of comparison, we have shown as the dashed (red) line on the same
plot the spectrum corresponding in the BCB, with the model given
in~\cite{ungarelli2001}. These two spectra will be used for some
computations in the sequel of this letter.

Using equations (\ref{eq:rotation}-\ref{eq:Sgw}), it is now possible
to deduce the correlation function $E_\gw$ now obtained after a
double averaging over the gravitational environment and over the
measurement time
\begin{equation}
E^{\A\B}_\gw=-\overline{\langle\cos\left(2(\Theta-\alpha)\right\rangle}
\;,
\end{equation}
where the symbols $\langle\;\rangle$ and $\bar{\quad}$ represent
respectively a trace over the stochastic gravitational waves
background and an averaging over the measurement time $T$. In a
linearized treatment, the variable $\alpha$ has a gaussian
distribution so that the correlation function $E^{\A\B}_\gw$ can be
rewritten
\begin{eqnarray}  \label{eq:reductionE}
&&E^{\A\B}_\gw=\mathcal{C}\,
\cos\left(2\Theta-2\overline{\langle\alpha\rangle}\right) \;, \\
&&\mathcal{C}=\overline{\langle\cos\left(2\delta\alpha\right)\rangle}
= \exp\left(-2\Delta\alpha^2\right) \;, \nonumber \\
&&\delta\alpha=\alpha-\overline{\langle\alpha\rangle} \;,\quad
\Delta\alpha^2= \overline{\langle\left(\delta\alpha\right)^2
\rangle} \;. \nonumber
\end{eqnarray}
The interaction with gravitational waves leads to two different
effects~: the first one is a mean rotation angle
$\overline{\langle\alpha\rangle}$ which, as already discussed, can
be compensated by rotating the polarizers. In contrast, the second
effect is a net reduction of the correlation function due to the
fluctuations $\delta\alpha$ which is an exponential function of the
variance $\Delta\alpha^2$ of these fluctuations.

For the CGWB, which corresponds to quantum fluctuations, the
variance may then be written as the following integral over
frequencies
\begin{equation}  \label{eq:variancecgwb}
\Delta\alpha^2_\cgwb = \int_0^\infty\frac{\mathrm{d}\omega}{2\pi}\,
S_\gw[\omega]\mathcal{A}[ \omega]\;.
\end{equation}
The dimensionless apparatus function $\mathcal{A}$ characterizes the
response of the experiment to gravitational waves at frequency
$\omega$. It depends on the geometry of the experiment, and in
particular on the time of flight $\tau$. Its calculation is done in
a similar way to that of similar functions appearing in the study of
decoherence in interferometers~\cite{gwdecoh} or of sensitivity to
gravitational waves of clock synchronization~\cite{reynaud2008}. At
the end of this calculation, it may be written
\begin{eqnarray}
&&\mathcal{A}[\omega] = \frac{5}{8}\int_{-1}^1\frac{\mathrm{d}\mu}{2}%
\,|\beta_\ra-\beta_\la|^2\;, \\
&&\beta_\ra[\omega,\mu]= (1-\mu)\left(e^{-i\omega\tau(1+\mu)}-1\right)
\nonumber \\
&&\beta_\la[\omega,\mu]= (1+\mu)\left(e^{-i\omega\tau(1-\mu)}-1\right)\;.
\nonumber
\end{eqnarray}
The amplitudes $\beta_\lra$ measure the sensitivity to the
gravitational mode at frequency $\omega$ and wavevector with
$\mu=ck^3/\omega$ the direction of $k^3$ ($\mu$ is defined for the
photon $\rightarrow$, and replaced by $-\mu$ for the photon
$\leftarrow$).

\begin{figure}[h!]
\onefigure{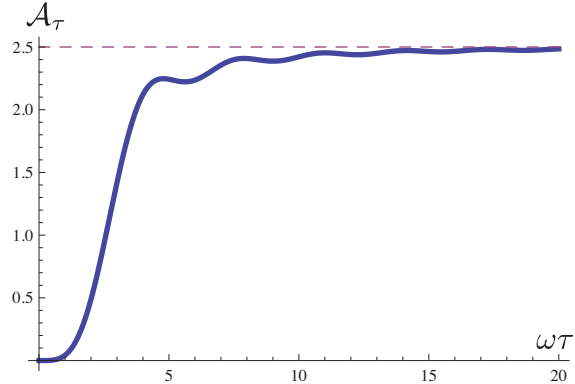} \caption{Apparatus function $\mathcal{A}_\tau$, drawn
versus $(\omega\tau)$.} \label{fig:A}
\end{figure}

After straightforward integrations, the function $\mathcal{A}$ is
finally written as
\begin{equation}  \label{eq:exactA}
\mathcal{A}[\omega]=\frac{5}{16} \left(8-\frac{24+6\cos(2\omega
\tau)}{(\omega \tau)^2}+\frac{15\sin(2\omega \tau)}{(\omega
\tau)^3}\right)\;.
\end{equation}
The result is drawn on Fig.~\ref{fig:A} as a function of $\omega
\tau$. It tends to the constant $5/2$ at the limit of high
frequencies and behaves as $(\omega\tau)^4/21$ at the limit of low
frequencies.
For the purpose of comparing the two effects, we give also the
variance for the BCB, which corresponds to classical fluctuations
and leads to a slightly different expression
\begin{eqnarray}  \label{eq:variancebcb}
&&\Delta\alpha^2_\bcb = \int_0^\infty\frac{\mathrm{d}\omega}{2\pi}\,
S_\gw[\omega]\mathcal{A}[\omega]\mathcal{F}[\omega] \\
&&\mathcal{F}[\omega]= 1 - \left(\protect\mathrm{sinc}\frac{\omega
T}2\right) ^2\;,\quad \protect\mathrm{sinc}(x)=\frac{\sin x}x \;.
\nonumber
\end{eqnarray}

The two variances (\ref{eq:variancecgwb}) and (\ref{eq:variancebcb})
are integrals of the gravitational waves background over frequencies
$\omega$ larger than the inverse of one of the typical time involved
in the correlation measurement. For the CGWB, this time is just the
time of flight $\tau$ of the photons from the source to the
detectors. For the BCB, this time is the longer one of the time of
flight $\tau$ and the measurement time $T$. The effect associated
with $\tau$ may be considered as intrinsic, as it cannot be reduced
for a given geometry. In contrast, the effect associated with $T$
can in principle be limited by having this measurement time shorter
than $\tau$.


\section{Discussion and conclusions}


We now come to the discussion of the decrease of EPR quantum
correlations, \textit{i.e.} the reduction of the maximal violation
of the CSHS inequality (\ref{eq:CHSH}), due to the effect of the
CGWB. For completeness, we also present results for the BCB.

From eq.(\ref{eq:reductionE}), we deduce the maximal value
$S_{\mathrm{max}}$ of the parameter $S$, after optimal angles have
been chosen,
\begin{equation}  \label{eq:reductionS}
S_{\mathrm{max}}= \mathcal{C}\,2\sqrt{2}=2\sqrt{2}\,
\exp\left(-2\Delta\alpha^2\right) \;.
\end{equation}
Following the analogy between the EPR correlation function and
interferences, it turns out that the reduction of the maximal
violation of Bell-CSHS inequality is analogous to the decoherence
effect already studied for interferometers~\cite{gwdecoh}. The main
difference is that the reduction is determined by the variance of an
angle $\alpha$ whereas it was determined by the variance of a phase
for interferometers.

\begin{figure}[h!]
\onefigure{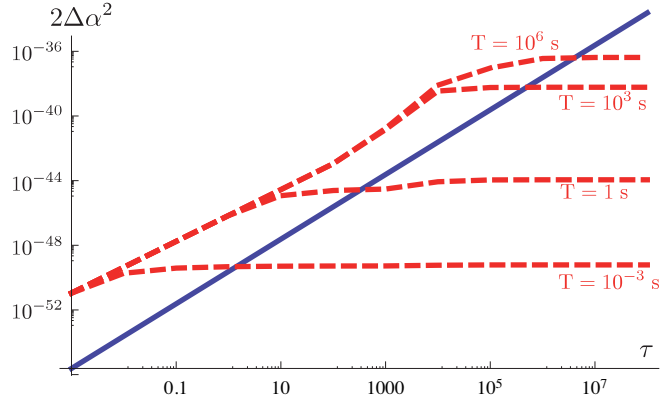} \caption{Variances $2\Delta\alpha^2$
appearing in the exponential factor in eq.(\ref{eq:reductionS})
plotted as functions of the time of flight $\tau$~: the full (blue)
line corresponds to the effect (\ref{eq:variancecgwb}) of the CGWB
($\Omega_\gw=10^{-14})$, and the dashed (red) line to the effect
(\ref{eq:variancebcb}) of the BCB ((model given
in~\cite{ungarelli2001}))~; in the latter case, the various curves
correspond to different values of the measurement time
$T=10^{-3},\,1,\,10^{3}$ and $10^6$s.} \label{fig:variances}
\end{figure}

We show on Fig.\ref{fig:variances} the results of the evaluation of
the quantity $2\Delta\alpha^2$ which appears in the exponential
factor in eq.(\ref{eq:reductionS}). The full (blue) line corresponds
to the effect (\ref{eq:variancecgwb}) of the CGWB
($\Omega_\gw=10^{-14})$, and the dashed (red) line to the effect
(\ref{eq:variancebcb}) of the BCB ((model given
in~\cite{ungarelli2001})). All curves are drawn as functions of the
time of flight $\tau$ varying from $10^{-3}$s - typical time of
flight from a station on Earth to the ISS - to $10^6$s - time of
flight corresponding to the scale of the solar system ($\approx2000$
astronomical units). For the case of the BCB, the various curves
correspond to different values of the measurement time $T$ varying
also from $10^{-3}$s to $10^6$s. As expected all variances are
increased when the time of exposition ($\tau$ or $T$) to the
gravitational wave backgrounds are increased. The effect of the BCB
is larger than that of the CGWB for not too large times of
exposition, which just reflects the shapes of the corresponding
spectra which were plotted on Fig.\ref{fig:spectra}.

We see on Fig.\ref{fig:variances} that the variance $\Delta\alpha^2$
remains extremely small for any scale accessible to experiments in
the solar system. This entails that the quantum correlations coded
in polarization entangled pairs of photons should survive the
exposition to gravitational wave backgrounds in any presently
foreseeable experiment. In fact, this conclusion can be made even
more general through an argument which sheds light on an unexpected
deep connection between the discussion of large scale EPR
correlations and cosmology.

\begin{figure}[h!]
\onefigure{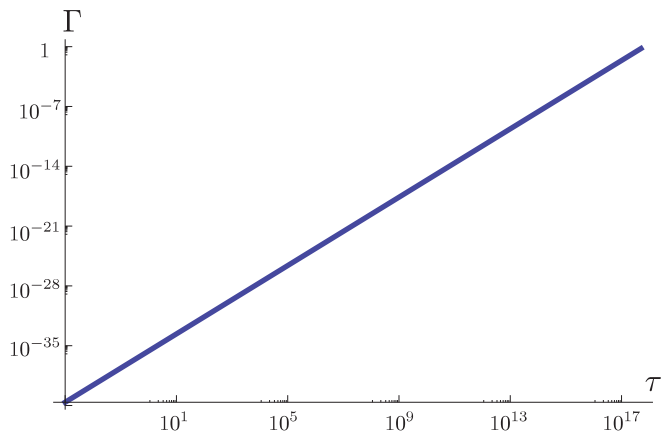}
\caption{Dimensionless factor $\Gamma$ as a function of the time of flight $%
\protect\tau$ on log-log scales; the full (blue) line corresponds to
the CGWB, while the dotted and dashed (red) lines corresponds to the
BCB with $T\ll\tau$ and $T=1000$s respectively.  }
\label{fig:relicnoise}
\end{figure}

In order to write down this argument, we may rewrite the reduction
factor $\mathcal{C}$ appearing in
eqs.(\ref{eq:reductionE},\ref{eq:reductionS}) as follows, under the
assumption of a constant value for $\Omega_\gw$,
\begin{eqnarray}
&&\mathcal{C}=e^{-\Gamma \Omega_\gw} \;,\quad \Gamma= \frac{24
H_0^2}{5}\, \int_0^\infty \frac{\dd\omega}{\omega^3} \,
\mathcal{A}[\omega] \;. \label{eq:reductioncosmic}
\end{eqnarray}
When the form (\ref{eq:exactA}) of $\mathcal{A}$ is used, the
integration leads to the closed expression
\begin{eqnarray}  \label{eq:Gamma}
&&\Gamma = \frac{3}{5} (H_0\tau)^2 \;.
\end{eqnarray}
The resulting $\Gamma(\tau)$ is the straight line appearing on the
log-log plot of Fig.\ref{fig:relicnoise}, which we have drawn for
$\tau$ varying up to the Hubble time $\tau\approx 10^{18}$s.

Fig.\ref{fig:relicnoise} shows that the number $\Gamma$ is extremely
small at all scales small with respect to the Hubble scale. In fact,
$\Gamma$ increases as the square of the time scale of the experiment
and it would approach unity only at the largest cosmological scales.
When evaluating the reduction factor $\mathcal{C}$ in
(\ref{eq:reductioncosmic}), the small number $\Gamma$ has first to
be multiplied by a second small number $\Omega_\gw$ and then
exponentiated. As a consequence, the reduction has to remain
negligible in any experiment. A different conclusion could
only be obtained at the largest cosmological scales, if the energy
of gravitational waves would be a dominant constituent of the
content of the Universe.

As a conclusion, we have shown that the quantum correlations coded
in polarization should survive the exposition to gravitational wave
backgrounds. In particular, initial quantum correlation arising from
primordial cosmic processes are still present in current day
Universe even if their detection is a challenge. Note that these
conclusions have been obtained for gravitational wave backgrounds as
they are predicted by standard calculations, and they would of
course be affected if larger fluctuations were produced by physical
processes beyond the standard model. Note also that this conclusion
relies on a fascinating and unexpected connection between the
reduction of EPR quantum correlations at large scales and the
fundamental cosmic parameter $\Omega_\gw$.


\end{document}